\documentclass[doublespacing]{elsart}

\usepackage{epsfig}
\usepackage{bm}

\journal{Physics Letters B}

\begin{document}

\begin{frontmatter}
\title{Test experiment to search for a neutron EDM by
  the Laue diffraction method}

\author[1]{V.V. Fedorov},
\author[1]{E.G. Lapin},
\author[2]{E. Leli{\`e}vre-Berna},
\author[2]{V. Nesvizhevsky},
\author[2]{A. Petoukhov},
\author[1]{S.Yu. Semenikhin},
\author[2]{T. Soldner},
\author[2]{F. Tasset},
\author[1]{V.V. Voronin\corauthref{cor}}
\corauth[cor]{Corresponding author.}
\ead{vvv@pnpi.spb.ru}

\address[1]{Petersburg Nuclear Physics Institute, Gatchina, Leningrad district, 188300, Russia}

\address[2]{Institut Laue Langevin, 6 rue Jules Horowitz, BP 156,
  F-38042 Grenoble Cedex 9, France}

\begin{abstract}
A prototype experiment to measure the neutron electric dipole moment (nEDM)
by spin-rotation in a non-centrosymmetric crystal in Laue geometry
was carried out in order to investigate the statistical sensitivity
and systematic effects of the method. The statistical sensitivity to the
nEDM was about  $6\cdot 10^{-24}$\,e$\cdot $cm per day and can be
improved by one order of magnitude for the full scale setup.
Systematics was limited by the homogeneity of the magnetic field in
the crystal region and by a new kind of spin rotation effect.
We attribute this effect to a difference of the two Bloch waves amplitudes
in the crystal, which is caused by the presence of a small crystal
deformation due to a temperature gradient. In a revised scheme of the
experiment, this effect could be exploited for a purposeful
manipulation of the Bloch waves.
\end{abstract}

\end{frontmatter}

\section{Introduction}

The search for a finite nEDM is a prominent example
of the quest for new sources of CP violation and thus for
physics beyond the Standard Model. The most precise
experiments today were carried out using the Ramsey resonance
method and ultra-cold neutrons (UCNs) \cite{pnpiedm,edmlast}.
Further progress is presently limited by systematics \cite{pendlebury2004}
and the low density of UCNs available.

In general, the statistical sensitivity of an experiment to measure the nEDM
is determined by the product $E\tau \sqrt{N}$,
where $E$ is the value of the electric field, $\tau$ the duration of the
neutron interaction with the field and $N$ the number of the counted neutrons.
New projects to measure the nEDM with UCNs
aim to increase the UCN density and thus $N$ by orders of magnitude (see
\cite{golub2005} for a recent overview).
In contrast, experiments with crystals exploit the electric
field inside matter, which can, for some crystals, be orders of magnitude
above the electric field achievable in vacuum.

EDM experiments with absorbing crystals were pioneered by Shull and Nathans
\cite{shull1967}. The first one based on the interference of electromagnetic amplitude with the imaginary part of the nuclear one \cite{shull1963}. The first who paid attention to the presence of a
spin dependent term due to interference of nuclear and spin-orbit parts of scattering amplitude
in the neutron interaction with a  non-centrosymmetric non-absorptive crystal were Abov
with his colleagues \cite{Abov1966}. 
Spin-rotation in non-centrosymmetric crystals due to such interference
as a way to search for a nEDM was first discussed by Forte \cite{forte1983}. The corresponding spin-rotation effect due to spin-orbit interaction was experimentally tested by Forte and Zeyen \cite{ForteZeyen}.
A similar to \cite{forte1983}, but more detailed theory of neutron optical activity and
dichroism for diffraction in non-centrosymmetric crystals has been developed in \cite{Barysh}.
Authors of the works \cite{grav,dfield} have  shown and experimentally proved that the interference of nuclear and electromagnetic parts of the scattering amplitude leads to an existence of a constant strong electric field, acting on a neutron during all time of its movement in the noncentrosymmetric crystal. This field was measured first \cite{dfield} in the neutron Laue diffraction experiment, the measured value being coincided with the calculated one.

Recently a new method of a neutron EDM search was proposed \cite{dedm,polart,PhysB2001,dedm1}.
The value of the electric field in this method was determined
experimentally
for quartz to $E\approx 2\cdot10^8$\,V/cm \cite{dfield,dptfe}.
The interaction time of the neutron with the electric field is
shorter than in UCN experiments and can reach $\tau\approx 1-2$\,ms
\cite{tfjetpl,PhysB2001,LDM_sens}. The statistical sensitivity
of the method profits from the higher flux of the used cold neutrons,
compared to UCNs available today.
In a test experiment \cite{LDM_sens} we have determined the statistical 
sensitivity of the method and have found that with existing quartz
crystals and
the flux of the PF1B beam line \cite{haese2002} of the ILL one
can reach $\sim 6\cdot 10^{-25}$\,e$\cdot$cm
per day, about the value of the most sensitive published UCN experiments
\cite{pnpiedm,edmlast}. 

The aim of the experiment presented in this paper was
to confirm the statistical sensitivity with a prototype set-up
and to investigate systematic errors of the method.

\section{Laue diffraction method to search for a nEDM}

In non-centrosymmetric crystals the diffracted neutrons are moving
in two Bloch states exposed to opposite electric fields
\cite{dfield,dedm} because of the shift of the ``electric'' planes relative
to the ``nuclear'' ones\footnote{In our notation, ``nuclear'' or ``electric''
planes are determined by the positions of the maxima
of the corresponding periodic potential.}.
In the reference system moving together with the neutron the electric field
is seen as a magnetic one. The interaction (Schwinger interaction)
of the neutron spin with this effective magnetic field
$\bm{H}^{\rm S}_{\rm g}=\frac{1}{c}[\bm{E}_{\rm g}\times \bm{v}]$
results in a spin precession around $\bm{H}^{\rm S}_{\rm g}$.
The spin rotation angle for the two states is 
\cite{dedm}
\begin{equation}
\Delta\phi^{\rm S}_{1,2}=\pm\frac{2\mu H^{\rm S}_{\rm g}}{\hbar}\tau=
  \pm\mu_{\rm n}\frac{eE_{\rm g}L}{m_{\rm p}c^2},
 \label{DfiS}
\end{equation}
where $\tau=L/v_\parallel$ is the interaction time ($v_\parallel$ being the
component of the neutron velocity
parallel to the diffracting planes and $L$ the thickness of the crystal) and
$\mu=\mu_{\rm n} e\hbar/2m_{\rm p}c^2$ the neutron magnetic moment with 
$\mu_{\rm n}=-1.9$.
The signs $\pm$ refer to the Bloch states 1 and 2, respectively.
For a longitudinally polarized beam with the incident polarization $P_0$,
the final neutron polarization $P$ is longitudinal and given by
\cite{PhysB2001,dedm1}\footnote{This result is obtained by averaging
the Pendell\"{o}sung oscillations over the Bragg angles. The angular period
of these oscillations in the experiment was $\sim 10^{-5}$\,rad and the
angular divergence of the neutron beam $\sim 10^{-2}$\,rad.}:
\begin{equation}
    P = P_0  \cos \Delta\phi^{\rm S}_{1,2}=
        P_0 \cos \left(\frac{\mu_{\rm n} e E_{\rm g}  L}{m_{\rm p}  c^2}\right).
	\label{SchwingerPolarisation}
\end{equation}
It can be decreased down to zero by choosing
the crystal thickness $L_0$ such that $\Delta\phi^{\rm S}_{1,2}=\pm\pi/2$.
The calculation for the $(110)$-plane of $\alpha$-quartz
gives $L_0=3.5$\,cm.

We consider Bragg angles $\theta_{\rm B}$ close to $\pi/2$, i.e.
$\bm{v}\parallel\bm{P}_0$ is
almost parallel to $\bm{E}_{\rm g}$. For each of the two Bloch states, a component
of the polarization vector perpendicular to $\bm{E}_{\rm g}$ builds up in the
XY-plane due to the Schwinger precession
Eqs.~(\ref{DfiS},\ref{SchwingerPolarisation}). These components as well
as the electric field $\bm{E}_{\rm g}$ have opposite signs for the two Bloch
states. The interaction of a finite nEDM with the
electric field $\bm{E}_{\rm g}$ results in a precession of the
polarization vector around $\bm{E}_{\rm g}$, thus creating a component
$P^{\rm EDM}$ in Z direction, with the same sign for the two Bloch states.
$P^{\rm EDM}$ is given by
\cite{PhysB2001,dedm1}:
\begin{equation}
P^{\rm EDM}= \frac{4DE_{\rm g}L_0}{\pi \hbar {v_\|}}=\frac{4D}{\pi\mu}\cdot\frac{c}{v \cos
\theta_{\rm B}}
\propto \frac{1}{\pi/2-\theta_{\rm B}}.
 \label{Ph}
\end{equation}
Here $D$ is the nEDM. Rotating the crystal by the angle $2\theta_{\rm B}$ 
($\approx\pi$)
changes the sign of $\bm{E}_{\rm g}$ and thus that of $P^{\rm EDM}$. This
is the experimental signature of a nEDM.

\begin{figure}[tbhp]
	\centering
		\includegraphics[width=0.8\textwidth]{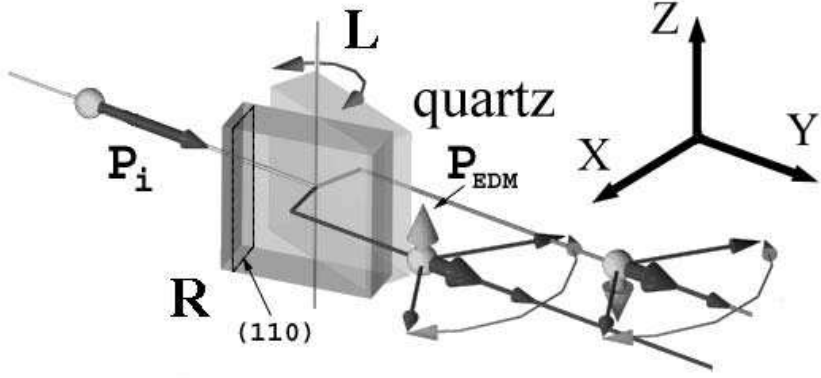}
    \caption{Principal scheme of the experiment for a nEDM search
    by the Laue diffraction method.
      The presence of a nEDM leads to a small Z-component of
      the polarization, which has opposite signs for the two crystal
      positions R and L.}
    \label{fig:Fig1}
\end{figure}

The principal scheme of the method is shown in Fig.~\ref{fig:Fig1}. We
assume that the environmental magnetic field is low enough as to avoid any
further spin rotation. For the two crystal positions right (R) and
left (L) with the same Bragg angle but with 
opposite directions of the electric field, the polarizations
$P^{\rm EDM}_{\rm R, L}$ have
opposite signs whereas a residual polarization has the same sign.
In the coordinates of Fig.~\ref{fig:Fig1}, $P^{\rm EDM}$ is given by
the difference of the Z components of the final polarization vectors
for the two crystal positions R and L. High precision of the crystal
rotation and ``zero'' magnetic field conditions
are necessary to exclude systematic errors and to select the EDM effect.

>From Eq.~(\ref{SchwingerPolarisation}) follows that the effect due to
the Schwinger interaction does not depend on neutron
properties such as the energy, the wavelength or the Bragg angle. It is
determined  by the property of the crystal and by the fundamental
constants only. For a given crystal it is the same for all Bragg
angles. In contrast, the EDM effect Eq.~(\ref{Ph}) depends 
on the Bragg angle and grows strongly for $\theta_{\rm B} \rightarrow \pi/2$.
Thus, carrying out the measurement for two different Bragg angles
gives an additional way to eliminate false effects related to the Schwinger
interaction.

\section{Experimental set-up}

In order to avoid systematic effects, the two crystal positions have to be
fully identical.
Any gradient of the residual magnetic field or a temperature gradient
over the crystal violates this requirement and, finally, can result
in a systematic offset on the final polarization which could mimic the
searched effect.

\begin{figure}[tbhp]
	\centering
		\includegraphics[width=1.0\textwidth]{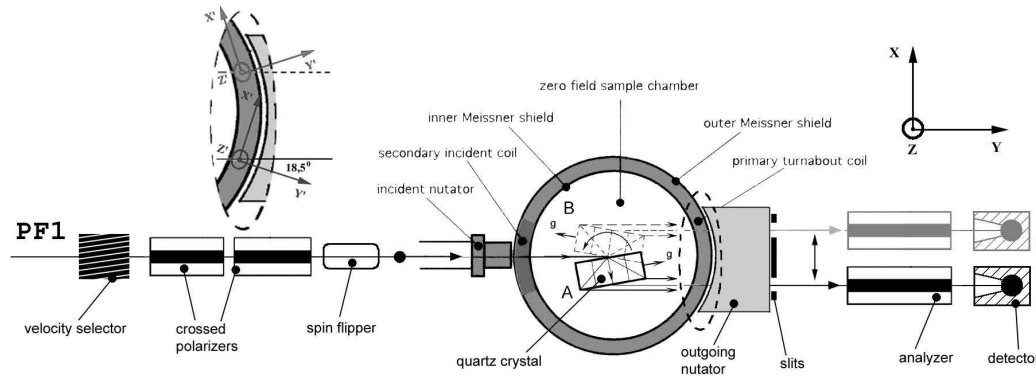}
\caption{General layout of the experiment.}
\label{fig:1}
\end{figure}

The experimental setup is shown in Fig.~\ref{fig:1}. 
It basically consists of the part responsible for the preparation of the
beam with the desired neutron wavelength and polarization (neutron velocity
selector, polarizer, flipper), the
part responsible for the zero-field condition on the sample and spherical
polarimetry (Cryopad with nutators \cite{Cryopad}), and 
the analyzer of the final polarization. 

The (110) plane ($d=2.456$\AA) of a quartz crystal was used in the experiment.
The size of the crystal was $140 \times 140\times 35$ mm$^3$ and its mosaic
$\sim 1$ angular second for the whole body of the crystal.

The experiment was carried out at the end position of the neutron
guide H53 (instrument PF1) of the ILL. The guide is connected to the horizontal
cold source and provides a capture flux of $3.5\cdot10^9$\,n/cm$^2$s.
The velocity selector served to preselect neutrons with a wavelength
of $(5.0\pm0.3)$\,{\AA} and thus reduced the background. To obtain a
homogeneous and wavelength-independent incident polarization, two super mirror
polarizers were used in crossed geometry \cite{kreuz2005}. With this
set-up, an incident polarization of about 99.5{\%} is expected
\cite{kreuz2005}.  The experimental values $AP=97.8${\%} for the direct
beam and $AP=(97.5\pm 0.2)${\%} for the diffracted beams were limited
by the analyzing power $A$ of the single super mirror analyzer. The
neutron spin could be flipped by a resonance flipper.

For spherical polarimetry, the so-called Cryopad-II was used.
It consists of two concentric cylindrical Meissner magnetic screens which
prevent the penetration of the external field inside the cavity at the
height of the beam. $\mu$-metal screens located above and below
Cryopad-II, together with the Meissner screens, assure a reduced magnetic
field in the sample chamber. Nutators and superconducting coils are used
to select the direction of the incident polarization vector and the
component of the outgoing polarization vector that is being analyzed.
The residual magnetic field inside the cavity was of the order of a few mG.
The cross-section of the beam at the sample was $0.8\times1.7$\,cm$^{2}$.

The cylindrical geometry of Cryopad leads to a discrepancy of the coordinate
systems between the crystal positions L and R (see Fig.~\ref{fig:1})
and limits the accuracy of the measurement of the polarization vector
at about $\pm $10\,mrad. We estimate that this factor and the residual
magnetic field result in a total directional uncertainty of the
polarimetry of about 20\,mrad.
This value agrees qualitatively with the typical precision obtained
with this second-generation Cryopad of 35\,mrad \cite{Cryopad}.

Although the superconducting screens are obviously thermically isolated,
the air temperature in the center of Cryopad was about
0.5\,K higher than that close to the Cryopad walls.

\begin{figure}[tbp]
	\centering
		\includegraphics[width=1.0\textwidth]{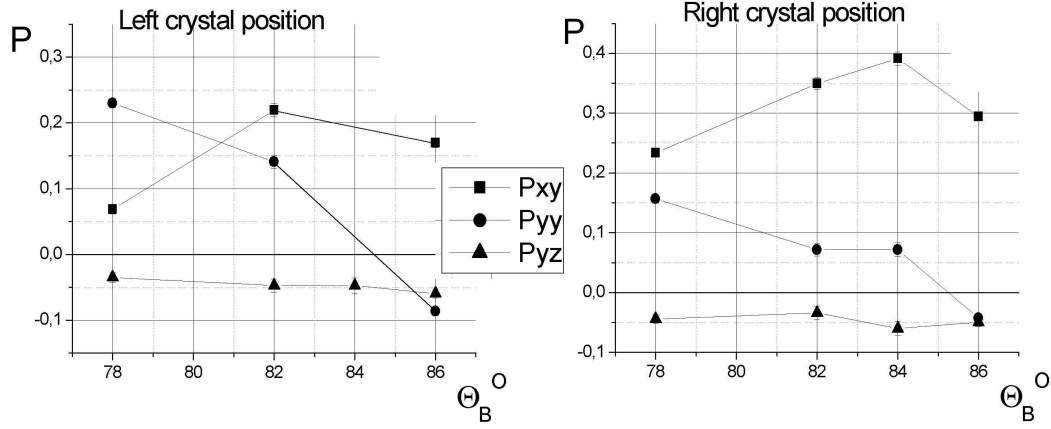}
\caption{Angular dependence of selected elements of the polarization
  matrix $P_{ij}$.}
\label{fig:3}       
\end{figure}

\section{Experimental results}

The intensity of the diffracted beam for the experimental geometry was
about 3-6\,n/s for the ``grey'' polarization direction. It coincides
with the result obtained in our previous experiment \cite{LDM_sens}. The
relatively small intensity was mainly limited by the small size of the
incident nutator restricting the beam cross-section to $0.8\times1.7$\,cm$^{2}$.
The corresponding statistical
sensitivity for the nEDM is $6\cdot 10^{-24}$ e$\cdot $cm
per day. This value can be improved by about one order of magnitude by
using a more intense neutron beam (for instance the
instrument PF1B \cite{haese2002} of the ILL) and increasing the sizes
of the beam and of the quartz crystal.
 
\begin{figure}[tbp]
	\centering
		\includegraphics[width=0.8\textwidth]{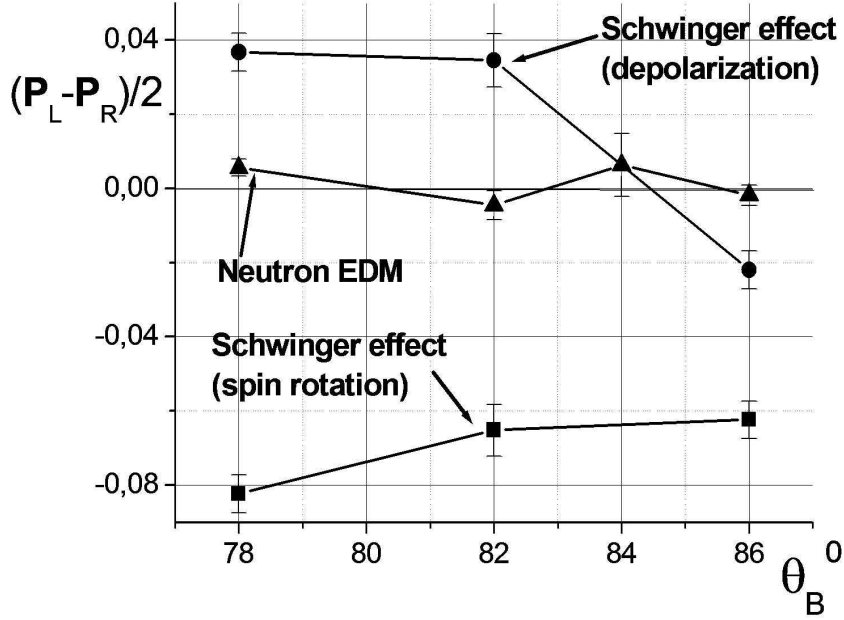}
\caption{Difference of selected elements of the polarization matrix $P_{ij}$
  between the two crystal positions R and L.}
\label{fig:4}       
\end{figure}

Examples of the measured angular dependences of the elements
of the polarization matrix $P_{ij}$
are shown in Fig.~\ref{fig:3}. Here, $i$ indicates the direction of the
incident polarization and $j$ the analyzed component of the final polarization.
The presence of a nEDM would lead to a
difference of $P_{yz}$ for the two crystal positions.
The observed difference (triangles in the Fig.~\ref{fig:4}) is consistent
with zero and can be used to estimate the nEDM
$D=(3.5\pm 1.6)\cdot 10^{-23}$\,e$\cdot$cm (only statistical error stated).
This value, obtained within a few hours of data collection,
is 20 times more precise than the result of the crystal 
diffraction experiment of Shull and Nathans \cite{shull1967}. 
We did not try to improve the statistical accuracy further
since we found large final polarizations $P_{xy}$, $P_{yy}$, see
Figs.~\ref{fig:3} and \ref{fig:4}, in contradiction to the expectations.

\section{Discussion}

To our understanding, the only reason for measuring
large elements $P_{xy}$, $P_{yy}$
of the polarization matrix is a difference between the two Bloch waves
amplitudes excited in the crystal.\footnote{Note that a nonzero value
of the matrix element $P_{xy}$ indicates a neutron spin rotation due to
spin-orbit interaction in Laue diffraction. Earlier, such an
effect was predicted only for absorbing crystals \cite{Barysh}.}

A difference of this type can be explained by the Borman effect (different
absorptions for the two Bloch waves). However, our estimations 
for the (110) plane of a quartz crystal show that this mechanism can
account for not more than 10\% of the observed effect. 

A deformation of the crystal could be another possible reason.
It is well known that the interaction of neutrons or X-rays
with elastically deformed crystals strongly differs from the undeformed
case \cite{L_R_Asy_JETP,Difr_Focus_JETP,grav}. In our experiment,
such a deformation could arise from the temperature gradient
of 0.5\,K in the Cryopad. The thermal expansion of quartz is
$\Delta L/L\approx 10^{-5}\Delta T/{\rm K}$. This has to be
compared with the Bragg width $\Delta\lambda_{\rm B}/\lambda\approx 10^{-5}$
of the (110) plane of the crystal.

The trajectories of neutrons in a deformed
crystal can be described by the so called ``Kato forces'' \cite{Kato_3}
that are determined by the value of the crystal deformation. 
For a constant gradient in the interplanar distance the neutron trajectories
inside the crystal are given, see \cite{grav,Kato_3}:
\begin{equation}
\frac{\partial ^2z}{\partial y^2} = \pm \frac{\tan^2\theta_{\rm B}}{m_0 }\pi g\xi, 
 \label{Fk}
\end{equation}
where $m_0\equiv 2d F_{\rm g}/V_{\rm c}$ is the so called ``Kato mass'' ($F_{\rm g}$ is the
neutron structure factor of the reflection, $V_{\rm c}$ the volume of the unit
cell, and $d$ the interplanar distance), $g=2\pi/d$ the reciprocal lattice
vector, and $\xi$ describes the crystal deformation ($d=d_0(1+\xi z)$).
The signs $\pm$ in Eq.~(\ref{Fk}) correspond to the two Bloch waves. 
Eq.~(\ref{Fk}) holds for crystal deformations small compared to the
Bragg width of the reflection.

\begin{figure}[tb]
	\centering
		\includegraphics[width=0.6\textwidth]{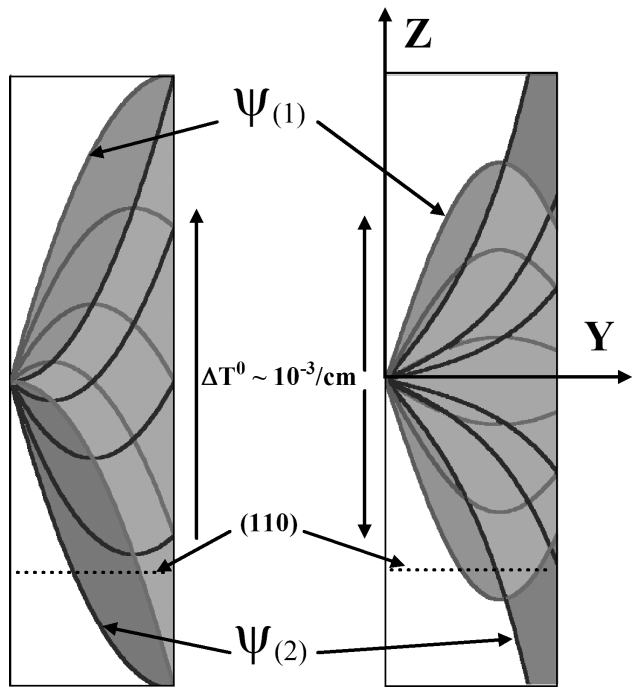}
	\caption{Examples of neutron trajectories in the deformed
	quartz crystal ((110) plane, $d=2.45$\AA, $\theta_{\rm B}=86^\circ$,
	crystal size $35\times 140$mm$^2$). The neutrons enter the crystal
	from the left side in the center. Left: Constant temperature
	gradient $d=d_0(1+\xi z)$. Right: The center of the crystal has a
	higher temperature than its sides, $d=d_0(1+\xi |z|)$. This
	was the case in the experiment.}
	\label{fig:Tr_1}
\end{figure}

The right part of Eq.~(\ref{Fk}) is proportional to the square of
$\tan \theta_{\rm B}$. In our experiment this factor was
rather high: $\tan^2 \theta_{\rm B}\sim 100\ldots 1000$
($\theta_{\rm B}\approx (84 - 87)^\circ$). This results in a
very high sensitivity of the neutron trajectories to even small deformations
of the crystal.

We calculated neutron trajectories in a deformed crystal
of $14\times3.5$\,cm$^2$ for the temperature gradient $\Delta T=10^{-3}$\,K/cm
and  $\theta_{\rm B}=86^\circ$ ($\tan^2\theta_{\rm B}=200$). The results are shown in
Fig.~\ref{fig:Tr_1}. The left plot corresponds to a constant temperature
gradient from one side of the crystal to the other and the right one to a
gradient from the center to both sides. The two cases behave 
differently: For the constant gradient the crystal deformations
do not result in different amplitudes for the two Bloch waves at the exit
surface. For the
gradient from the center to both sides, one Bloch wave ($\Psi(1)$ in
Fig.~\ref{fig:Tr_1}) is focused, the other one defocused. Consequently,
the amplitudes of the two Bloch waves at the exit surface of the crystal
are not equal.
The outgoing neutron wave is the superposition of the two Bloch
waves where the spin was rotated in opposite directions. Because of
their different amplitudes, the depolarization effect described by
Eqs.~(\ref{DfiS},\ref{SchwingerPolarisation}) is incomplete. For equal
amplitudes, the X components of the polarization vectors of the
two Bloch waves cancel everywhere (compare Fig.~\ref{fig:Fig1}). For
different amplitudes this is not the case and thus the total polarization
vector is rotated in the XY-plane.
In our model, it is the new spin rotation effect which is responsible
for the high values of the elements $P_{xy}$ and $P_{yy}$ observed in the
experiment.

It is important to point out that, although $P^{\rm EDM}$ is measured
via the Z-component of the final polarization vector, the surviving
polarization in the XY-plane can cause an offset to $P^{\rm EDM}$,
for example due to a residual magnetic field that turns the polarization
toward the Z direction.

On the other hand, the effect permits to manipulate the amplitudes of
the Bloch waves by applying a temperature gradient. For the inverse sign
of the gradient, for example, $\Psi(1)$ is defocused and $\Psi(2)$ focused.
Manipulating the sign of the temperature gradient is equivalent
to changing the sign of the effective electric field seen by the forward
diffracted neutrons. By a high temperature gradient one of the Bloch waves
can even be fully suppressed.

\begin{figure}[tb]
	\centering
		\includegraphics[width=0.8\textwidth]{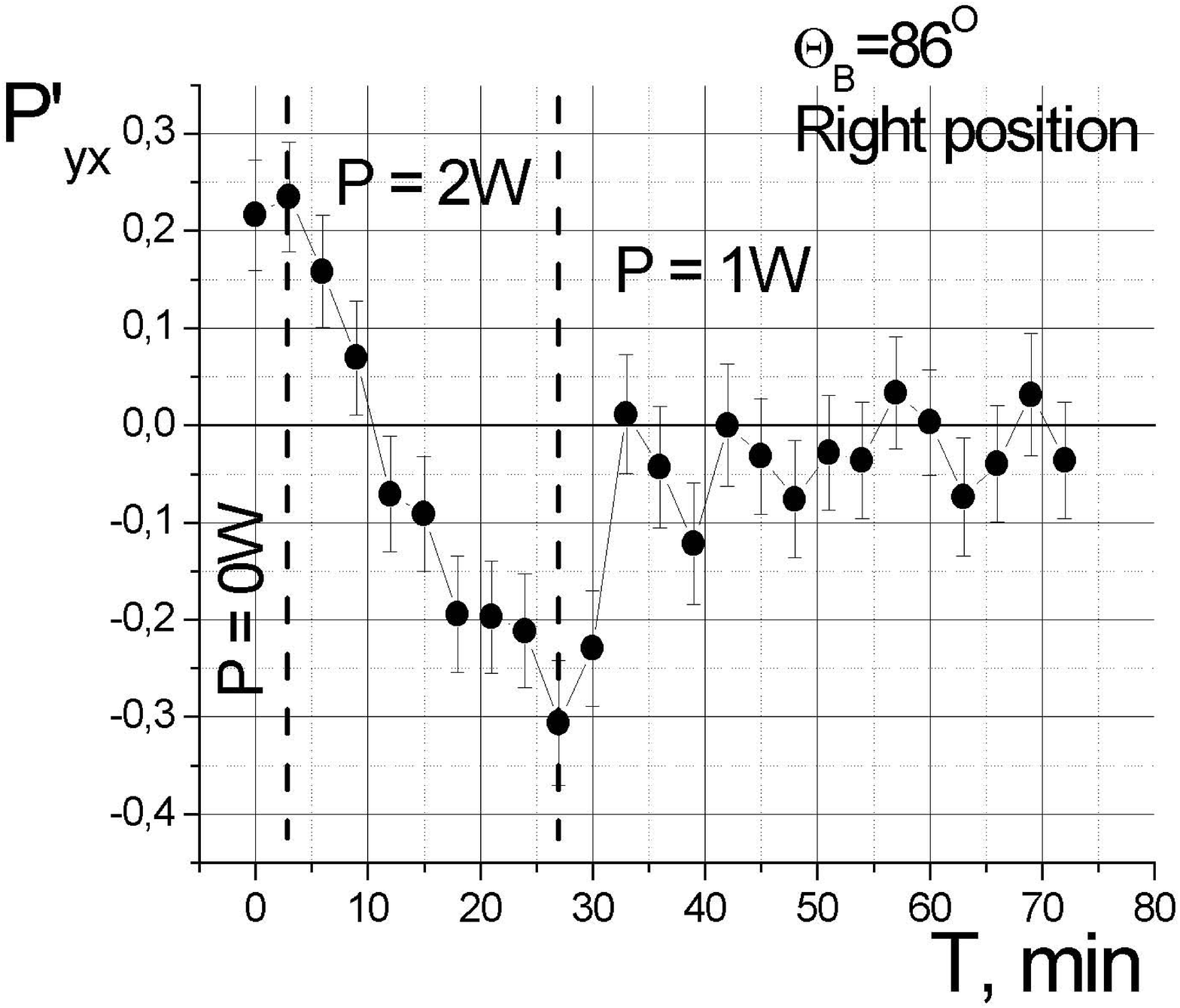}
\caption{Dependence of $P_{yx}$ on the heating power applied to one side of
  the crystal, for the crystal position R.}
\label{fig:2}       
\end{figure}

To test our explanation of the observed effect, we
intentionally introduced a temperature gradient along the crystal
by installing a small electric heater close to one surface of the
crystal and measured the spin rotation effect ($P_{yx}$)  for different
heating powers. The results are shown in Fig.~\ref{fig:2}:
Heating the crystal with an electric power of about 2\,W changed the
sign of $P_{yx}$. The process had a relaxation time of the order of 
half an hour. Reducing the power of the heater to 1\,W resulted in the
compensation of the spin-rotation effect for the used crystal position. 
These results are in agreement with our explanation.

\section{Conclusions}

A prototype experiment to measure the nEDM by Laue diffraction in
non-centrosymmetric crystals was carried out. 
This test allowed us to determine experimentally the statistical sensitivity
of the method and to investigate possible systematic errors.

The experiment confirmed our expectation
as regards the high statistical sensitivity: already the first run of
a few hours resulted in the value $D=(3.5\pm1.6)\cdot10^{-23}$\,e$\cdot$cm
which is 20 times more precise than the result of the previous experiment
based on crystal diffraction \cite{shull1967}. With a dedicated installation,
a statistical precision of the order of $6\cdot 10^{-25}$\,e$\cdot$cm per day
can be achieved, comparable to the present nEDM experiments with
UCNs and the Ramsey resonance method \cite{pnpiedm,edmlast}.

Beside the statistical precision the most important issue is the question
about systematic errors. The main source of systematic errors is a residual
magnetic field in the zero-field cavity where the crystal is installed. In our
installation we used an old version of Cryopad as cavity,
which has a relatively high residual magnetic
field of the order of a few mG. This field limits the EDM
experiment to a precision of about $10^{-24}$\,e$\cdot$cm. We believe that
it is possible
to build a dedicated spherical polarimeter based on the Cryopad
ideas to reach a precision of the EDM experiment of the order of a few 
times $10^{-26}$\,e$\cdot$cm. 

The performed experiment also allowed us to discover a new effect: for
forward diffraction of polarized neutrons by a non-centrosymmetric crystal
of quartz (110 plane) a large ($\approx\pi/4$) spin-rotation effect induced
by a temperature gradient was observed. We identified this
effect as a combination of the Schwinger interaction of the
neutron spin with the crystal interplanar electric field and 
the modification of the Bloch waves amplitudes by a 
temperature induced deformation of the crystal.

The consequences of this new spin-rotation effect
are twofold: on the one hand it imposes a limitation on the
proposed method to search for a nEDM and requires serious revision
of the method, on the other hand it can serve as a new tool to
manipulate the sign and the value of the crystal interplanar electric
field and, hence, the sign of the nEDM effect. These possibilities
open a new road to develop a nEDM experiment 
exploring the high interplanar field of non-centrosymmetric crystals.

\section*{Acknowledgments}

The authors would like to thank the personnel of the ILL reactor (Grenoble,
France), in particular E. Bourgeat-Lami and S. Pujol, for the technical
assistance in the experiment.
This work was supported by RFBR (grants No 05-02-16241-a,
03-02-17016-a) and INTAS.


\begin{thebibliography}{50}

\bibitem{pnpiedm} I.S.~Altarev, Yu.V.~Borisov, N.V.~Borovikova, et
al., Yad.Fiz. {\bf 59} (1996) 1204.

\bibitem{edmlast} P.G.~Harris, C.A.~Baker, K.~Green, et al.,
Phys. Rev. Lett. {\bf 82} (1999) 904.

\bibitem{pendlebury2004}
  J.M.~Pendlebury, W.~Heil, Yu.~Sobolev, P.G.~Harris, J.D.~Richardson, R.J.~Baskin, D.D.~Doyle, P.~Geltenbort, K.~Green, M.G.D.~van~der~Grinten, P.S.~Iaydjiev, S.N.~Ivanov, D.J.R.~May, and K.F.~Smith,
  Phys. Rev. A {\bf 70} (2004) 032102.

\bibitem{golub2005} R.~Golub and P.R.~Huffman,
  J. Res. Natl. Inst. Stand. Technol. {\bf 110} (2005) 169.

\bibitem{shull1967} C.G.~Shull and R.~Nathans, Phys. Rev. Lett.
{\bf 19} (1967) 384.

\bibitem{shull1963} C.G.~Shull,  Phys. Rev. Lett.  {\bf 10} (1963)  297.

\bibitem{Abov1966} Yu.G.~Abov, A.D.~Gulko, and P.A.~Krupchitsky,
  {\it Polarized Slow Neutrons}, (Atomizdat, Moscow, 1966)  256p. (in Russian).

\bibitem{forte1983} M.~Forte, J. Phys. G {\bf 9} (1983) 745.

\bibitem{ForteZeyen} M.~Forte and C.M.E.~Zeyen,
Nucl. Instr. Meth. \textbf{A284} (1989) 147.

\bibitem{Barysh}  V.G.~Baryshevskii and S.V.~Cherepitsa,
Phys. Stat. Sol. {\bf b128} (1985) 379; Izvestiya Vuzov SSSR, ser. fiz. {\bf 8 } (1985) 110 (in Russian).

\bibitem{grav} V.L.~Alexeev, E.G.~Lapin, E.K.~Leushkin, V.L.~Rumiantsev,
  O.I.~Sumbaev, and V.V.~Fedorov, JETP {\bf 94} (1988) 371.

\bibitem{dfield} V.L.~Alexeev, V.V.~Fedorov, E.G.~Lapin, et al.,
Nucl. Instr. Meth. A {\bf 284} (1989) 181; JETP {\bf 69} (1989) 1083.


\bibitem{dedm} V.V.~Fedorov, V.V.~Voronin, and E.G.~Lapin,
  J. Phys. G. {\bf 18} (1992) 1133.

\bibitem{polart} V.V.~Fedorov, V.V.~Voronin, E.G.~Lapin, and
  O.I.~Sumbaev, 
  Tech. Phys. Lett.  {\bf 21} (1995) 884;
  Physica {\bf B234--236} (1997) 8.

\bibitem{PhysB2001}
  V.V.~Fedorov, E.G.~Lapin, S.Yu.~Semenikhin, and V.V.~Voronin,
  Physica B {\bf 297} (2001) 293.

\bibitem{dedm1} V.V.~Fedorov and V.V.~Voronin,
  Nucl. Instr. Meth. B {\bf 201} (2003) 230.


\bibitem{dptfe}
  V.V.~Voronin,  E.G.~Lapin, S.Yu.~Semenikhin, and V.V.~Fedorov,
  JETP Lett. {\bf 72} (2000) 308.

\bibitem{tfjetpl}
  V.V.~Voronin, E.G.~Lapin, S.Yu.~Semenikhin, and V.V.~Fedorov,
  JETP Lett. {\bf 71} (2000) 76.

\bibitem{LDM_sens} V.V.~Fedorov, E.G.~Lapin, E.~Leli{\`e}vre-Berna,
  V.~Nesvizhevsky, A.~Petoukhov, S.Yu.~Semenikhin, T.~Soldner, F.~Tasset,
  and V.V.~Voronin, Nucl. Instr. Meth. B {\bf 227} (2004) 11.

\bibitem{haese2002} H.~H{\"a}se, A.~Kn{\"o}pfler, K.~Fiederer, U.~Schmidt, D.~Dubbers, and W.~Kaiser,
  Nucl. Instr. Meth. A {\bf 485} (2002) 453.

\bibitem{Cryopad} F.~Tasset, P.J.~Brown, E.~Leli{\`e}vre-Berna, T.~Roberts,
  S.~Pujol, J.~Allibon, and E.~Bourgeat-Lami, Physica B {\bf 267-268} (1999) 69.

\bibitem{kreuz2005} M.~Kreuz, V.~Nesvizhevsky, A.~Petoukhov, and T.~Soldner,
  Nucl. Instr. Meth. A \textbf{547} (2005) 583.

\bibitem{L_R_Asy_JETP} Yu.S.~Grushko, E.G.~Lapin, O.I.~Sumbaev, and
  A.V.~Tyunis, JETP {\bf 47} (1978) 1185.
 
\bibitem{Difr_Focus_JETP} O.I.~Sumbaev and E.G.~Lapin, JETP {\bf 51} (1980) 403.


\bibitem{Kato_3} N.~Kato, J. Phys. Soc. Japan {\bf 19} (1964) 971.
 
\end{thebibliography}
\end{document}